\documentclass[reqno,12pt]{article}
\usepackage{amsmath,amsfonts,amssymb,amsthm,amstext,amscd,eucal,xcolor}
\usepackage[all]{xy}
\usepackage{hyperref}
\usepackage{epsfig}
\usepackage{color}
\usepackage{graphicx}
\usepackage[active]{srcltx}
\makeatletter \@addtoreset{equation}{section}

\makeatletter\renewcommand\section{\@startsection {section}{1}{\z@}%
                                   {-3.5ex \@plus -1ex \@minus -.2ex}
                                   {2.3ex \@plus.2ex}%
                                   {\normalfont\large\bfseries}}
\renewcommand\subsection{\@startsection{subsection}{2}{\z@}%
                                     {-3.25ex\@plus -1ex \@minus -.2ex}%
                                     {1.5ex \@plus .2ex}%
                                     {\normalfont\bfseries}}

\newcommand{\be}{\begin{equation}}
\newcommand{\ee}{\end{equation}}
\newcommand{\bea}{\begin{eqnarray}}
\newcommand{\eea}{\end{eqnarray}}
\newcommand{\bse}{\begin{subequations}}
\newcommand{\ese}{\end{subequations}}
\newcommand{\bi}{\begin{itemize}}
\newcommand{\ei}{\end{itemize}}

\newcommand{\beq}{\begin{eqnarray}}
\newcommand{\eeq}{\end{eqnarray}}

\newcommand{\nn}{\nonumber}

\def\s2s1{S$^2\times$S$^1$ }
\def\Label#1{\label{#1}%
  \smash{\hbox to0pt{\raise1ex\hbox{\tiny[#1]}\hss}}}
\def\noLabels{\let\Label=\label}
\def\nobbibitem{\let\bbibitem=\bibitem}

\begin{document}
\baselineskip 18pt%
\begin{titlepage}
\vspace*{1mm}%
\hfill%
\hfill \vbox{
    \halign{#\hfil         \cr
arXiv:yymm.nnnn\cr
         } 
      }  
\begin{center}
{\Large{\textbf{Interacting Dark Energy: Dynamical
System Analysis}}} \vspace*{8mm}

{  Hanif Golchin\footnote{h.golchin@.uk.ac.ir}$^{,a}$, Sara Jamali\footnote{sara.jamali@stu.um.ac.ir}$^{,b}$
,Esmaeil Ebrahimi\footnote{eebrahimi@uk.ac.ir}$^{,a,c}$}\\
\vspace*{0.4cm}
{$^a$ \it Faculty of Physics, Shahid Bahonar University, PO Box 76175, Kerman, Iran} \\
{$^b$ \it Department of Physics, Ferdowsi University of Mashhad, PO Box 1436 Mashhad, Iran}\\
{$^c$\it Research Institute for Astronomy and Astrophysics of
Maragha (RIAAM), Maragha, Iran} \\
\vspace*{1.0cm}
\end{center}

\begin{abstract}
We investigate the impacts of interaction between dark matter and dark
energy in the context of two dark energy models, holographic and ghost dark energy. In fact, using the
dynamical system analysis, we obtain the cosmological consequence of several interactions, considering all relevant component of universe, i.e. matter (Dark and luminous), radiation and dark energy. Studying the phase space for all interactions in detail,  we show the existence of unstable matter dominated and stable dark energy dominated phases. We also show that linear interactions suffer from the absence of standard radiation dominated epoch.
Interestingly, this failure resolved
by adding the non-linear interactions to the models. We find an upper bound for the value of the coupling constant of the interaction between dark matter and dark energy as $b<0.57$\, in the case of holographic model, and $b<0.61$ in the case of ghost dark energy model, to result in a cosmological viable matter dominated epoch. More specifically, this bound is vital to satisfy instability and deceleration of matter dominated epoch.
\end{abstract}

\end{titlepage}
\addtocontents{toc}{\protect\setcounter{tocdepth}{2}}
\tableofcontents

\section{Introduction}

Observations from supernovae type Ia (SNIa) revealed that the
universe is experiencing a phase of acceleration
\cite{riess,perl}. In other observations from SNIa and also from
cosmic microwave background radiation (CMBR) the issue is
confirmed \cite{spergel,cole,kowalski,spergel2,komatsu}. The
source for such an unexpected acceleration in general relativity
(GR) framework is the so-called ``dark energy (DE)". Simplest
candidate for DE is cosmological constant which
provide a vacuum energy background responsible for recent
acceleration. However observations indicates a small variations in
the equation of state (EoS) parameter, $w$, of the DE component.
Following such observations, some people turned to the dynamical
models of DE which have a variable EoS. This approach to DE is
widely discussed in the literature. For instance in \cite{pad},
the authors tried a tachyonic scalar field which the scalar field
play the role of  DE . A k-essence model of DE is considered by
Scherrer \cite{scherrer}. In \cite{arbey}, A unified model for
explaining DE and dark matter (DM) is presented and the authors introduced a complex scalar filed responsible for galactic DM
and the cosmic acceleration.

Ghost dark energy (GDE) is one of interesting models which is
based on the Veneziano ghost field in theory of Quantum ChromoDynamics (QCD) \cite{urban,ohta}. In
\cite{CaiGhost,urban}, authors showed that contribution of the
Veneziano ghost field is capable to derive an acceleration in the
cosmic background. In fact considering the value of energy scale in QCD as $\Lambda_{\rm QCD}\sim 100 MeV$ and $H
\sim 10^{-33}eV$, The GDE model alleviates the fine tuning problem\cite{ohta}. One can find different features of
GDE in
\cite{Nojiri:2005sr,Capozziello:2005pa,sheykhi1,sheykhi2,sheykhi3,sheykhi4,chao1,chao2,chao3}.

Another DE model which attracted considerable interest is the
so-called ``Holographic Dark Energy" (HDE). The base of this model
is the holographic principle which asserts that the number of
degrees of freedom for a physical system is related to its bounding
area rather to its volume \cite{Suss1}. In the light of this point, Li
proposed DE density as $\rho_D \leq3c^2m^2_p/L^2$ \cite{Li}. Here, $c^2$ is
a constant, $L$ denotes the IR cutoff radius and
$m^2_p =(8\pi G)^{-1}$. HDE is on of the most studied models of DE
and is capable to explain many features of cosmic evolutions. One
can refer to \cite{Huang,Hsu,HDE,Setare,Seta1,Setare1} for more
details. This models is also observationally constrained
\cite{Xin,Feng}.

Beside the DE component in the universe there exist a dark matter
component. One important task is to verify if these two component
can interact with each other. Theoretically there is not any
reason against their interaction and basically they can exchange
energy which affects the cosmic evolution. There also exist evidences that interacting models make better
agreement with observations \cite{interact1,oli}. Interacting
models of DE and DM entered the literature with \cite{wett}. In the
absence of an underlying theory of DE and DM, the form of
interaction term is a matter of choice. The simplest choice for
interaction term can be the linear combination of the form $Q\propto \rho_D+\rho_m$. However
other choices are also studied. For example in \cite{jian}, the
authors found that a model with a productive form of interaction
term ($Q\propto \rho_D \rho_m$) leads a good consistency with
observations. In the light of all mentioned above, it is well
motivated to consider non-linear interaction terms and study
their impacts on the cosmic evolution. 

Our main aim in this paper is to investigate imprints of non-linear interaction terms on DE models.
We study the evolution of DE models, by means of dynamical system analysis which is a powerful method and it has been frequently used in cosmology and astrophysics \cite{ketab,cope,amend,sari,Jamali:2016zww,Capozziello:2013bma,Landim:2015poa,Landim:2015uda}.

The paper is outlined as follows: in the next section we
briefly review GDE and HDE models in the flat universe with interaction between DM and DE, and present the necessary equations which we use them in the following. In section \ref{dsgde} (\ref{dshde}) we study the evolution of GDE (HDE) with different types of interaction terms, by using the dynamical system analysis. We summarize our results in the conclusion section.

\section{Interacting GDE and HDE models in flat universe}\label{sec-2}
Considering a flat universe filled with radiation, matter\footnote{By matter we mean all kind of matter, dark and luminous.} and dark energy. The first Friedmann equation is
\begin{eqnarray}\label{Fried}
H^2=\frac{8\pi G}{3} \left(\rho_r+ \rho_m+\rho_D \right),
\end{eqnarray}
where $\rho_r$, $\rho_D$ and $\rho_m$ are the energy densities of
radiation, dark energy and  matter,
respectively. Let us introduce the fractional energy density
parameters as
\be\label{Omega} \Omega_r=\frac{\rho_r}{\rho_{cr}}= \frac{8\pi G
\rho_r}{3 H^2},\qquad \Omega_m=\frac{\rho_m}{\rho_{cr}}=
\frac{8\pi G \rho_m}{3 H^2},\qquad
\Omega_D=\frac{\rho_D}{\rho_{cr}}=\frac{8\pi G \rho_D}{3H^2}, \ee
where $\rho_{cr}=3H^2/(8\pi G)$\,. 

According to the GDE model \cite{caighost2}, the energy density
of the dark energy defined as
\be \label{eden}
\rho_D=\alpha\,H+\beta H^2\,,
\ee
where $\alpha$ and $\beta$ are constants with dimension $(mass)^3$ and $(mass)^2$ respectively. In the GDE model this mass is $\Lambda_{\rm QCD}$, the mass scale of QCD, so the value of $\alpha$ ($\beta$) is of the order $\Lambda_{\rm QCD}^3$ ($\Lambda_{\rm QCD}^2$). Noting that $\Lambda_{\rm QCD}\sim 100 MeV$ and $H
\sim 10^{-33}eV$, the energy density of the GDE obtains as $\rho_D \sim 10^{-10}\,eV^4$. This is of the same order of observed value of the dark energy, so the GDE model does not face the fine tuning problem\cite{ohta}. Note also that in the present time, $\beta H^2$ term in (\ref{eden}) is subleading however, as it has been discussed in \cite{caighost2,Maggiore:2011hw} and in the following, this term could be notable in the early evolution of the universe.

The holographic principle also leads to another model to dark energy. In fact, Cohen et al. have shown \cite{Coh} that a short distance (UV) cutoff in quantum field theory could be related to a long distance (IR) cutoff $L$ due to the limit sets by black hole formation. In the other words, supposing that quantum zero-point energy  density $\rho_D$ is due to a UV cutoff, then the total energy in the region of size $L$ should not exceed the mass of a black hole of the same size, it means that $L^3 \rho_D \leq L M_p^2$\,. Now the longest $L$ is the one that saturating inequality and the HDE density takes the form
\be \label{hdeden}
\rho_D=3c^2M_p^2L^{-2}
\ee
where $c$ is a dimensionless constant and $M_p=\frac{1}{\sqrt{8\pi G}}$ is the reduced Planck mass. The IR cutoff $L$ can be chosen in different manner. If we set $L$ as the size of universe (the Hubble length), then the resulting energy density is the same order of the present day dark energy but this choice leads to wrong value for the EoS parameter. Instead, by choosing the future event horizon defined as
\be \label{feh}
R_h=a \int_t^{\infty}\frac{dt}{a}=a \int_a^{\infty}\frac{da}{a^2H}\,,
\ee
the correct EoS parameter could be obtained \cite{Li}.

Taking  interaction $Q$ between dark matter and the dark energy components to account,
the continuity equations read
\bea \label{conti1}
&& \dot\rho_r+4H\rho_r=0\,, \qquad  \dot\rho_m+3H\rho_m=Q\,,\\
&& \dot\rho_D+3H\rho_D(1+w_D)=-\,Q\,.\label{conti2} 
\eea
In the above, $Q>0$ denotes transition of energy content in the universe from DE to DM component and vice versa. The sum of equations in (\ref{conti1}), (\ref{conti2}) gives the total energy conservation in the universe as $\dot \rho+3H(\rho_{tot}+p_{eff})=0$\,, where the total equation of state can be written as
\be \label{teos}
w_{eff}=\frac{p_{eff}}{\rho_{tot}}=-1-\frac{2\dot H}{3H^2}\,,
\ee

Considering an interaction between DE and DM, the natural question is what will be the form of the
interaction term? 
Because of the unknown nature of DM and DE, there is no answer to this question based on particle physics theories, however such an interaction should be phenomenologically relevant. Remember also that ghost dark energy is a model which tries to answer the acceleration of the universe without any additional fields or degrees of freedom, So we choose the interaction terms in a manner that respect this outstanding feature of the model.

At the simplest level, the form of
interaction term is linearly related to $\rho_D$, $\rho_m$ or
the total energy density, $\rho_{tot}$. It is also logical to consider an interaction term proportional to $\rho_m \rho_D$. Choosing this product form means that the transfer rate of DE to DM (or vice versa) is negligible when $\rho_m, \rho_D \to 0$.
It has been shown that such a product coupling is consistent with
observations \cite{jian}.  On the other hand, Arevalo et.al introduced \cite{arevalo}
 several form of non-linear interaction term and
discussed their impacts on cosmic dynamics. Interactions in \cite{arevalo} can be accounted as a subset of  a general form for interaction terms as
\be \label{genint}
Q=3b^2H\rho_m^{\gamma}\rho_D^{\delta}\rho_{tot}^{\sigma}\,,
\ee
where $\gamma$, $\delta$ and $\sigma$ are integer numbers and it is obvious from the dimensional analysis that they satisfy $\gamma+\delta+\sigma=1$\,.
In the following sections we will investigate the evolution of GDE and HDE models accompanied by interaction terms in the form (\ref{genint}).

\section{The evolution of interacting GDE} \label{dsgde}
In this section we study the
evolution of the GDE model as a dynamical system. We start from GDE without interaction
between dark matter and dark energy, then add the linear
interaction and finally we study the impact of non-linear
interactions on the evolution of the model.  To investigate
the evolution of the model from early times, we consider the
contribution of radiation component in the energy contents of the
universe. 
By differentiating (\ref{eden}) and noting (\ref{conti2}) one finds
that
\be \frac{3H\rho_D(1+w_D)+Q}{H \rho_D}=-\frac{\dot
H}{H^2}\,\frac{\alpha+2\beta H}{\alpha+\beta H}\,, \ee
on the other
hand, differentiating the Friedmann equation (\ref{Fried}) and
noting (\ref{conti1}), (\ref{conti2}) one finds
\be \frac{\dot
H}{H^2}=-\frac12\,\big[3\Omega_m+4\Omega_r+3\Omega_D\,(1+w_D)\big]\,,
\ee
now, doing some calculations, the EoS parameter for dark energy the
deceleration parameter can be found as
\bea \label{wdqds} &&\!\!\!\!\!w_D=\frac{\alpha  \big[2
\text{$\Omega $}_q\!-\!H \text{$\Omega $}_D \left(3
(\text{$\Omega$}_D\! +\!\text{$\Omega $}_m\!-\!2)\!+\!4
\text{$\Omega $}_r\right)\big]\! +\!2 \beta H \big[\text{$\Omega
$}_q\!-\!H \text{$\Omega $}_D \left(3 (\text{$\Omega $}_D\!
+\text{$\Omega $}_m\!-\!1)+\!4 \text{$\Omega $}_r\right)\big]}{3 H
\text{$\Omega $}_D \big[\alpha (\text{$\Omega $}_D-2)
+2 \beta  H (\text{$\Omega $}_D-1)\big]}\,,\nn \\
&& \hspace{4cm}
q=-1+\bigg[\frac32\Omega_m+2\Omega_r+\frac32\Omega_D\,(1+w_D)\bigg]\,,
\eea
where $\Omega_q=\frac{8\pi G Q}{3H^2}$\,. In order to apply the phase space analysis, using the Friedmann
equation (\ref{Fried}), we introduce the dimensionless dynamical
variables $x$, $y$ and parameter $m$ as
\be \label{xydef} 
x^2= \Omega_m\,, \qquad \quad y^2=\frac{8\pi G \alpha}{3H}\,, \quad m^2=\frac{8\pi G \beta}{3}\,; \quad y^2+m^2=\Omega_D\,,
\ee
consequently the radiation density parameter reads as $\Omega_r=1-m^2-x^2-y^2$. Note that due to the constant value of $\beta$, there is a constant part $m^2$ in the dark energy density parameter $\Omega_D$. 
At the present time, considering $H\sim 10^{-33} eV$,\, $m^2$ is negligible
however, this subleading part might be significant in the early evolution of the universe as the early time dark energy \cite{Maggiore:2011hw}. In fact, it has been shown \cite{caighost2} that $m^2$ could have a fraction energy density about 10$\%$ in the early universe so in the following we will refer $m^2$ as early dark energy (EDE).
In the other word, the parameter $m$ always satisfies $0<m^2\leq 0.1$\,. We also refer to the $y^2$ part of the $\Omega_D$ as late dark energy (LDE).

After some algebraic manipulations the general form of the dynamical equations, which are generalization of the Friedmann equations, take the
form \bea \label{deqs}
x'&=&\frac{x^2 \left(2 m^2+2 x^2+5 y^2-2\right)+f(x,y) \left(2 m^2+2 x^2+y^2-2\right)}{2 x \left(2 m^2+y^2-2\right)}\,,\nn\\
y'&=&\frac{y\big[\left(4 m^2+x^2+4 y^2-4\right)+ f(x,y)\big]}{2\left(2 m^2+y^2-2\right)}\,,
\eea
 where the prime denotes derivative
with respect to $\ln a$ and $f(x,y)=\frac{\Omega_q}{H}$.  It seems that (\ref{deqs}) blow up when {\it i}) $y^2=2-2m^2$ and {\it ii}) $x=0$\,. However considering that $0< y < 1$ and $0<m^2\leq 0.1$, the condition ({\it i}) never satisfied and so $\left(2 m^2+y^2-2\right)$ does not vanish. To investigate the condition ({\it ii}), one should analyze (\ref{deqs}) in the presence of interaction term $f(x,y)$. In the following we consider GDE with six type of interactions and find the fixed points of the dynamical equations. We show that for three interactions, the dynamical equations are smooth every where. In one case, the dynamical equations are smooth conditionally at $x=0$  and in two other cases $x'$ diverges at $x=0$. We explain the physical meaning of divergency in these two cases. Note also that in general, the phase space of the interacting model is multi dimensional and the dynamical equations depends on several variables, however remembering that $m$ is a constant and also considering the form of interaction (\ref{genint}) between the DM and DE, there is only two dynamical variables $x$, $y$. In the other words the phase space of the model is two dimensional. Using the introduced dynamical variables, $q$, $\omega_{DE}$ and $\omega_{eff}$ can be found as
\bea \label{wqw}
w_D&=&\frac{2 m^4+m^2 \left(2 x^2+3 y^2-2\right)+y^2 \left(x^2+y^2+2\right)+2 f(x,y)}{3 \left(m^2+y^2\right) \left(2 m^2+y^2-2\right)}\,,\\
q&=&\frac{2 m^2+x^2+3 y^2-2+f(x,y)}{2 m^2+y^2-2}\,, \hspace{0.7cm}
 w_{eff}=\frac{2 \left(m^2\!+x^2-\!1\right)+5y^2\!+2 f(x,y)}{3 \left(2 m^2+y^2-2\right)}\,.\nn
\eea
 In the following, we consider the ghost dark energy model with interactions mentioned in the previous section and discuss the results in the context of  dynamical systems.

\vspace{3mm}
{\bf I) \,  The non-interacting case $Q=0$.}\\
In the absence of interaction between dark matter and dark energy ($f(x,y)=0$),
the dynamical equations (\ref{deqs}) are smooth in the range of variables. They have three acceptable fixed points:

$\bullet$ \, $P_{1}$: $(x=0,\, y=0)$. \, 
In this case, matter and LDE do not contribute in the energy content of the universe; it means that this fixed point describes early stages in the evolution of the universe. Remembering the comments after (\ref{xydef}), one can deduce that the universe is in a radiation/EDE scaling phase, where EDE fractional energy density is around $0.1\, (m\sim0.3)$, this yields the ratio $\Omega_r/\Omega_{EDE}\sim 9$ so this is in fact a radiation dominated era. In this case the eigenvalues of the stability matrix $\lambda_1=\frac 12$, $\lambda_2=1$  shows the instability of this phase. Moreover, using (\ref{wqw}), one can also 
finds $w_{eff}=\frac 13$ and $q=1$ which represent the decelerating expansion at this epoch.
 
$\bullet$ \, $P_{2}$: $(x=\sqrt{1-m^2},\, y=0)$. \, Considering the comments after (\ref{xydef}), one concludes that $P_2$ corresponds to a
matter/EDE scaling phase of the universe (similar to the previous case since $\Omega_{EDE}=m^2\sim 0.1$\,, this phase is actually matter dominated). In this case $w_{eff}=0$ and $q=\frac 12$ and the
eigenvalues of the stability matrix 
$\lambda_1=-1$, $\lambda_2=\frac 34$ shows the instability of matter dominated epoch for a
non-interacting universe.

$\bullet$ \, $P_{3}$: $(x=0,\, y=\sqrt{1-m^2})$. \, Remember that $m$ takes a very small value at the late times so the critical point $P_3$ demonstrates LDE dominated universe where  $q=-1$ and similar to the $\Lambda$CDM model 
$w_{eff}=w_D=-1$\,. This is a stable dark energy dominated epoch due to the eigenvalues of the stability matrix
$\lambda_1=-\frac 32$, $\lambda_2=-4$\,.

\vspace{3mm}
{\bf II) \,  The case $Q=3b^2 H \rho_{tot}$.}\\
For this linear interaction, where
$\rho_{tot}=\rho_r+\rho_m+\rho_D$, one finds that $f(x,y)=3b^2$\,. The dynamical equations (\ref{deqs}) in this case takes to the form
\be
x'\!=\!\frac{x^2\!\! \left(2 m^2\!\!+\!2 x^2\!\!+5 y^2\!\!-\!2\right)\!+\!3 b^2\! \left(2 m^2\!\!+2 x^2\!\!+y^2\!\!-\!2\right)}{2 x \left(2 m^2+y^2-2\right)}, \quad y'\!=\!\frac{y\! \left(3 b^2\!\!+\!4 m^2\!+\!x^2\!\!+4 y^2\!\!-\!4\right)}{2\! \left(2 m^2\!+y^2\!-\!2\right)},
\ee
Investigation of the above equations results in the following critical points:

$\bullet$ \, $P_1$: $(x=\sqrt{1-m^2},\, y=0)$. This point corresponds to the matter/EDE scaling phase of the universe. Remembering (\ref{wqw}), one finds that $w_{eff}=\frac{-b^2}{1-m^2}$, $q=\frac
12 \left(1-\frac{3b^2}{1-m^2}\right)$. As we expect the deceleration parameter is
positive, since the coupling $b$ has a small positive value as $b^2<\frac {1-m^2}{3}$\, where $m^2\leq 0.1$ \footnote{Although the behavior of $w_{eff}$ is not standard in this case, but the universe is decelerating since $w_{eff}>-\frac 13$.}. 
Keeping in mind this points, one confirms the instability of the
matter dominated epoch, considering the eigenvalues of the
stability matrix which are $\lambda_1=\frac 34(1-b^2)$ and $\lambda_2=-1-3b^2$\,.

$\bullet$ \, $P_2$: $(x=b,\, y=\sqrt{1-b^2-m^2})$. \, Noting that $m^2$ is small at the late times, this critical point
describes a LDE/matter scaling solution. As one already
knows $b$ has small value, hence the contribution of dark energy
is dominant. In this case the eigenvalues are
$\lambda_1=-3+\frac{6 b^2}{b^2+1}$
and $\lambda_2=-4$\,. Note that $\lambda_1$ takes negative values Since $b^2<\frac13$, so the dark
energy dominated epoch is stable. For this point one finds the
deceleration and EoS parameters as $q=-1$ and
$w_D=\frac{1}{b^2-1}$\,, which shows the phantom crossing behavior in this era due to the small value of $b$\,.

Although the late time behavior of the system in this case is accepted, the dynamical equation $x'$ diverges on $x=0$ line. In fact this line is excluded from the phase space and the linear interaction does not provide $x=0=y$ fixed point which means the absence of radiation dominated epoch. Therefor, in the context of GDE model, the linear interaction is not cosmologically accepted.

\begin{figure}[tt]
\begin{picture}(0,0)(0,0)
\put(114,-214){\footnotesize Fig (1.a)} \put(340,-214){\footnotesize Fig (1.b)}
\end{picture}
\center
\!\!\!\!\includegraphics[height=70mm,width=73mm,angle=0]{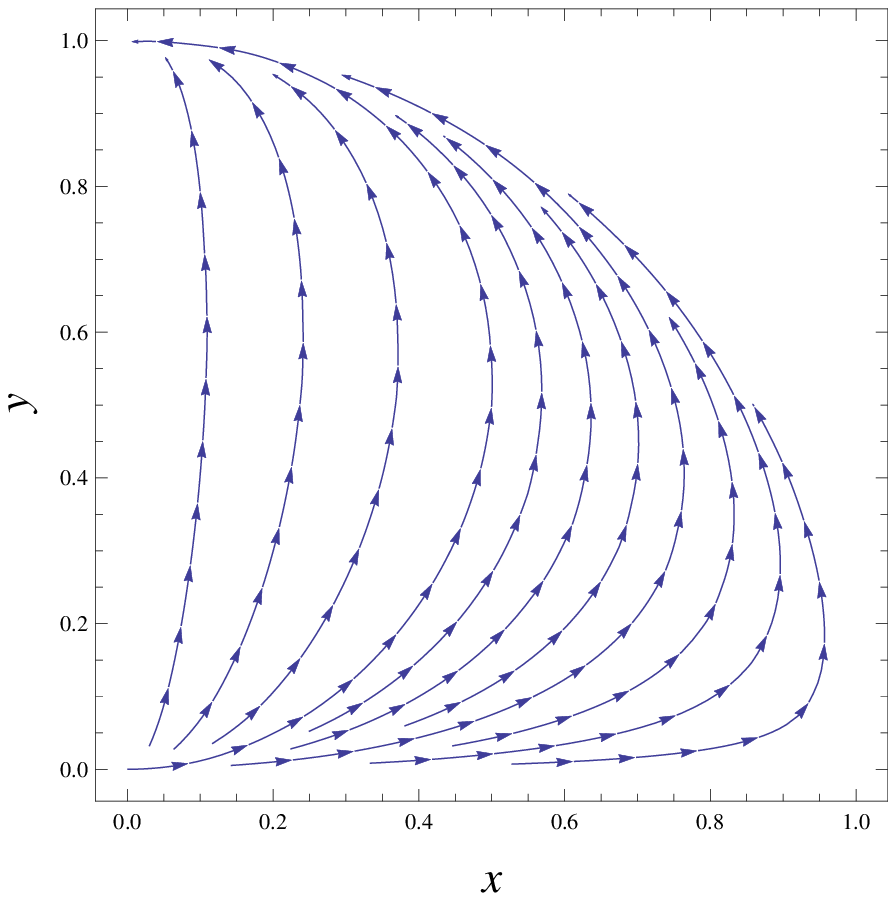}\,\,\,\,\,\, \includegraphics[height=70mm,width=73mm,angle=0]{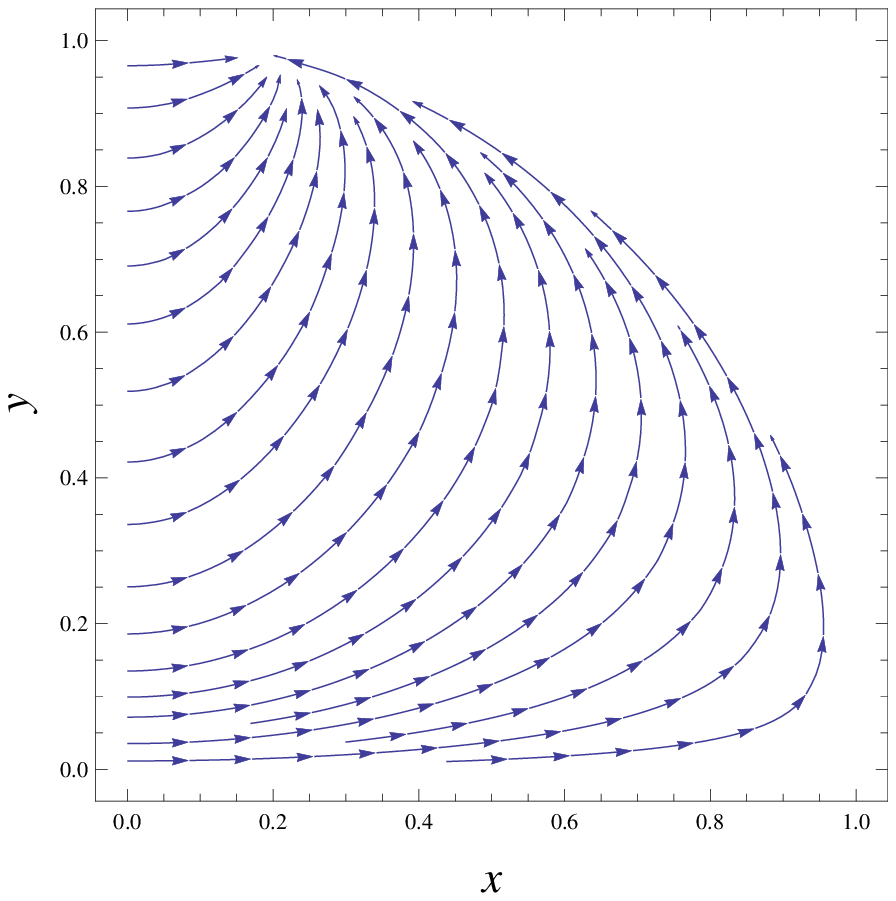}
\vspace{3mm} \caption{{\footnotesize The evolution of phase space
by choosing $b=0.2$\,. (1.a) corresponds to the non-interacting
GDE model which has unstable radiation dominated $(x=0,y=0)$, unstable
matter/EDE scaling $(x\sim1,y\sim0)$ and stable dark energy dominated $(x=0,y=1)$
fixed points. (1.b) corresponds to the linear interacting GDE.
This model shows unstable matter/EDE scaling and stable dark
energy-matter scaling behavior. Note that this model suffers from
the absence of radiation dominated fixed point ; this problem
resolved by adding some kind of non-linear interaction terms as we will see in
the following.}} \label{fig4}
\end{figure}

Figure (\ref{fig4}.a) shows the phase plane for non-interacting
case. It is easy to see that all arrows end at the point ($x=0,
y=1$)\footnote{Since $m^2$ is so small at present and future, we have written $P_3$ as ($x=0, y=1$).} which corresponds to a dark energy dominated universe.
True cosmological paths start from unstable radiation/EDE scaling phase ($x=y=0$), passing through unstable matter/EDE scaling era
($x=\sqrt{1-m^2}, y=0$) and end at the stable dark energy
dominated points ($x=0, y=1$). Phase plane of linear interacting
case is depicted in figure (\ref{fig4}.b). As explained above, the $x=0$ line in this figure is excluded from the phase space and the GDE with linear interaction term suffers from the absence of radiation
dominated epoch in the early times. Note also that the late time attractor lies at a point that the value of $x$ does not vanish. It shows that
adding linear interaction between matter and dark energy leads to
a scaling solution at late time. Considering $\Omega_D=y^2+m^2=1-b^2$,
to obtain a dark energy dominated universe, coupling $b$ of the
interaction must be small.

\vspace{3mm}
{\bf III) \,  The non-linear interaction $Q=3b^2 H\frac{\rho_D \rho_m}{\rho_{tot}}$.}\\
 Considering (\ref{Omega}) and (\ref{xydef}), one can finds that $f(x,y)=3b^2x^2 \big(y^2+m^2\big)$\,, so the dynamical equations (\ref{deqs}) for this non-linear interaction are
\bea
x'&=&\frac{x \left[y^2 \left(b^2 (9 m^2\!+6 x^2-6)+5\right)+2 \left(3 b^2 m^2\!+1\right) \left(m^2\!+x^2-1\right)+3 b^2 y^4\right]}{2 \left(2 m^2+y^2-2\right)}\,,\nn\\
y'&=&\frac{y \left[m^2 \left(3 b^2 x^2+4\right)+3 b^2 x^2 y^2+x^2+4 y^2-4\right]}{2 \left(2 m^2+y^2-2\right)}\,.
\eea
It is obvious that the above equations are smooth on  ranges of the variables. By solving the dynamical equations in this case, we found three physically acceptable critical points as

$\bullet$ \, $P_1$: $(x=0,\, y=0)$. \, Similar to the non-interacting case, this point corresponds to an unstable radiation/EDE scaling phase in the early stages of the universe with $m\sim 0.3$\, (note that $\frac{\Omega_r}{\Omega_{EDE}}\sim 9$, so one can call it radiation dominated phase). By using (\ref{wqw}) one finds that $q=1$ and $w_{eff}=\frac 13$\,. 
The instability of this phase can be deduced from the positive eigenvalues of the stability matrix, $\lambda_1= 1$ and
$\lambda_2=\frac 12(1+3b^2 m^2)$\,.

$\bullet$ \, $P_2$: $(x=\sqrt{1-m^2},\, y=0)$. \, This matter/EDE scaling era (where $\Omega_{EDE}=m^2\sim0.1$) is
unstable due to having one positive eigenvalue of the stability
matrix. In fact,  one finds that $\lambda_1=-1$ and
$\lambda_2=\frac34(1-b^2m^2)$ where $\lambda_2$ takes positive values, since $b^2m^2<<1$. In this case one obtains $w_{eff}=-b^2m^2>-\frac 13$ and $q=\frac 12 (1-3b^2m^2)>0$ which shows the deceleration of matter/EDE era.

$\bullet$ \, $P_3$: $(x=0,\, y=\sqrt{1-m^2})$. \, Remember that $m^2$ is very small at the late times hence, this point corresponds to dark energy dominated phase.  Eigenvalues for this critical point are $\lambda_1=-\frac32 (1-b^2)$ and
$\lambda_2=-4$\,. Therefor, stability of $P_3$ puts constraint $b<1$ on the coupling constant of interactions between DM and DE. The values of
$w_{eff}=w_D=-1$ and $q=-1$ in this case, are the same as standard
$\Lambda$CDM dominated solutions.

Note that adding the non-linear interaction in this case leads to
appearance of expected radiation dominated epoch in the early
times. This unstable epoch is absent in the case of linear
interaction in the above, and in \cite{GarciaSalcedo:2012dn}\,.

\vspace{3mm}
{\bf IV) \,  The non-linear interaction $Q=3b^2 H\frac{\rho_m^2}{\rho_{tot}}$.}\\
 In this case replacing the above interaction in (\ref{deqs}) and noting (\ref{Omega}), (\ref{xydef}) one
can find  dynamical equations as
\be
x'=\frac{x}{2} \!\left[\frac{2(3 b^2 x^4\!-\!4 m^2\!\!+x^2\!+4)}{2 m^2+y^2-2}+3 b^2 x^2\!\!+5\right],\quad y'=\frac{y (3 b^2 x^4\!\!+4 m^2\!\!+x^2\!\!+4 y^2\!-4)}{2 \left(2 m^2+y^2-2\right)},
\ee
which are smooth in the variables range. There are three physically accepted fixed points: 

$\bullet$ \, $P_1$: $(x=0,\, y=0)$. \, This point demonstrates an
instable radiation/EDE scaling phase (where $\Omega_r=0.9$) in the early universe. In this epoch, using (\ref{wqw}), one finds $w_{eff}=\frac 13$ and $q=1$ . The instability of this decelerating epoch,
confirmed by the eigenvalues $\lambda_1=1/2$ and
$\lambda_2=\frac12$\,.

$\bullet$ \, $P_2$: $(x=\sqrt{1-m^2},\, y=0)$. \, A matter/EDE scaling phase with $\Omega_{EDE}=m^2\sim 0.1$ is described by this critical point which is unstable due to eigenvalues
$\lambda_1=-(1+3b^2)$ and
$\lambda_2=\frac34 (1-b^2)$\,. In this case, one obtains that $w_{eff}=-b^2(1-m^2)$ and $q=\frac12 (1-3b^2(1-m^2))$. Note that to find a decelerating  matter dominated phase, the EoS parameter should satisfy $w_{eff}>-\frac 13$ and $q$ must be positive. This puts an upper bound on the coupling of interaction between DM and DE as  $b<\frac{1}{\sqrt{3(1-m^2)}}=0.61$\,. Inserting this bound, it is also clear that $\lambda_2>0$ and the instability of this matter dominated phase is confirmed.

$\bullet$ \, $P_3$: $(x=0,\, y=\sqrt{1-m^2})$. \, This point shows a late time
attractor, which means a stable dark energy dominated era. The
stability is deduced by negative eigenvalues $\lambda_1=-3/2$ and
$\lambda_2=-4$\,. One also can read from (\ref{wqw}) that
$w_{eff}=w_D=-1$ and $q=-1$ which is the same as a $\Lambda$-dominated solutions.

In this case similar to the case III, an unstable
radiation dominated epoch appears at the early times, due to the non-linear
interaction. Such an important epoch is absent in the case of
linear interaction II and in \cite{GarciaSalcedo:2012dn}\,, and
this shows the necessity of non-linear interaction terms for the
GDE model.

\begin{figure}[tt]
\begin{picture}(0,0)(0,0)
\put(5,129){\footnotesize L}
\put(5,120){$\bullet$}
\put(40,-14){\footnotesize Fig (2.a)} \put(199,-14){\footnotesize Fig (2.b)}
\put(379,-14){\footnotesize Fig (2.c)}
\end{picture}
\!\!\!\!\!\!\!\!\includegraphics[height=45mm,width=45mm,angle=0]{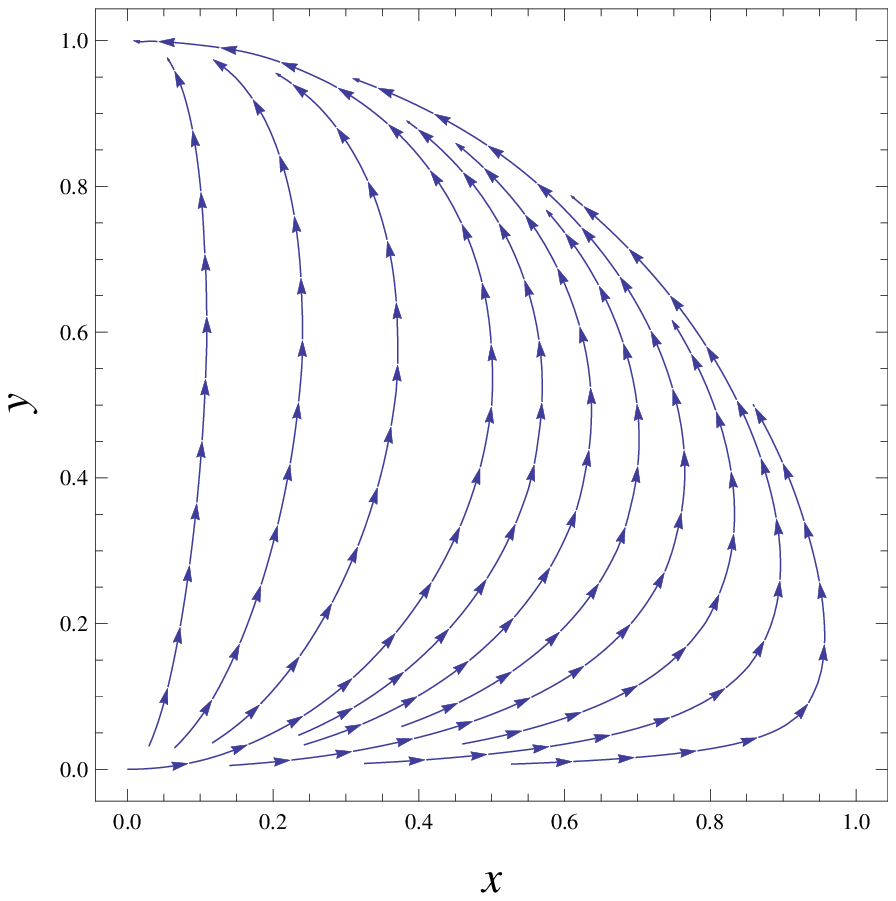}\,\,\,\includegraphics[height=45mm,width=65mm,angle=0]{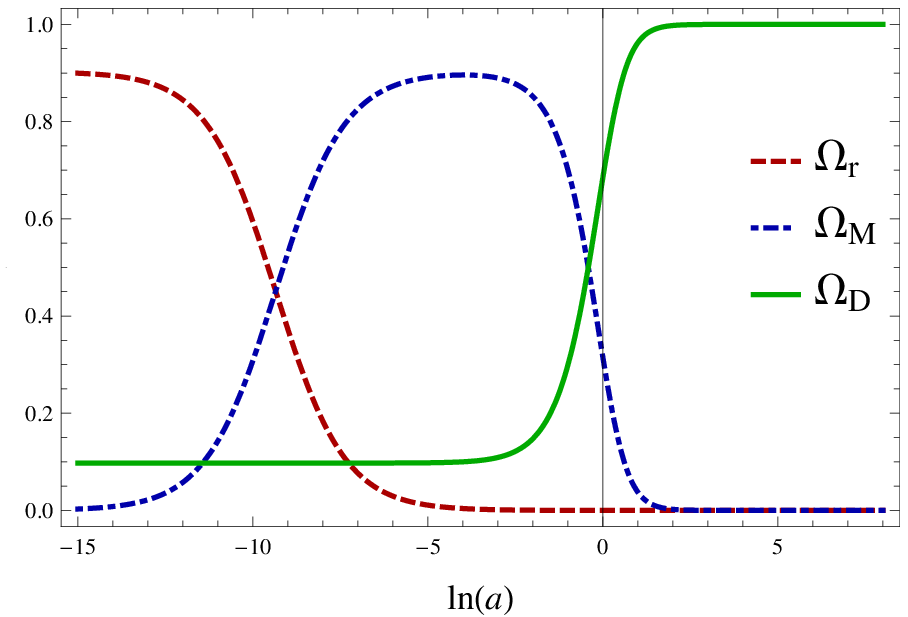}\,\,\,\includegraphics[height=45mm,width=60mm,angle=0]{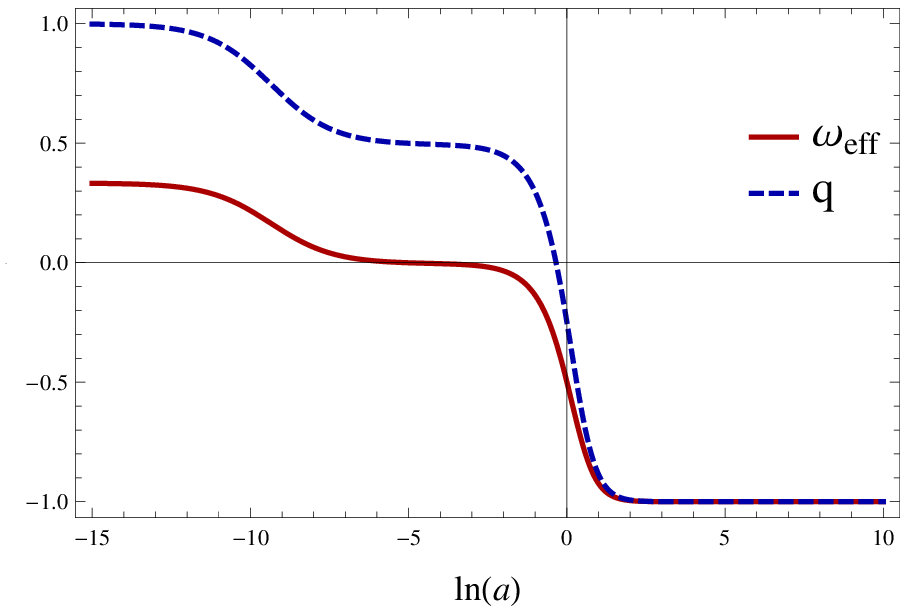}
\vspace{2mm} \caption{{\footnotesize (2.a) Shows the evolution of phase space of GDE with non-linear interactions III and IV, choosing $b=0.2$\,.
Note that adding these non-linear terms, the fixed point $(0,0)$ appears in the phase plane. This
fixed point which corresponds to unstable radiation dominated ($\Omega_r\sim 0.9$)
epoch, was absent in the case of GDE with linear interaction
term. \,(2.b) Presents the cosmic evolution of $\Omega_i$.  the initial conditions are chosen at present as $\Omega_D \approx 0.7$ and $\Omega_m \approx 0.3$ which is consistent to the observation.\, (2.c) Shows the evolution of EoS and deceleration parameters.}} \label{fig5}
\end{figure}

Phase space of the GDE with interactions III and IV are very similar to each other. Figure (\ref{fig5}.a) shows this phase plane. The phase plane contains many different initial conditions which are not necessarily physically accepted, but the phase plane demonstrate the instability of initial phase of universe and the existence of  a stable fixed point $L:(x=0, y\approx1)$ which corresponds to late time dark energy dominated universe. Specifically, true
cosmological paths are those that start from unstable radiation/EDE
scaling point ($x=y=0$), passing through unstable matter/EDE
 era ($x=\sqrt{0.9}, y\approx 0$), reaching $x^2\approx 0.3$, $y^2\approx 0.7$ at present and finally end at the
stable dark energy dominated points $L$\,, as shown in (\ref{fig5}.a). 
Although  GDE model with linear interaction suffers from the absence of radiation
dominated era in the early universe, this problem
resolved if one add the non-linear interactions in the form III
and IV to the GDE model.

We have depicted the evolution of fractional density parameters for the GDE model with interactions III, IV in figure (\ref{fig5}.b) where $\ln a=-\ln(1+z)$\,. By tunning the initial conditions, we found the expected values $\Omega_D\approx 0.7$ and $\Omega_m\approx 0.3$ at present time where $\ln a=0=z$. It is clear that in the past times, there is a constant early dark energy with $\Omega_{EDE}\sim 0.1$ and at large $z$ (early times), the model is radiation dominated ($\Omega_r\sim 0.9$)\,. There is also a transient matter dominated phase and finally the model reaches to a stable dark energy dominated phase. The evolution of EoS and deceleration parameters $w_{eff},\, q$\, is also plotted in figure (\ref{fig5}.c)\,.

\vspace{3mm}
{\bf V) \,  The non-linear interaction $Q=3b^2 H \frac{\rho_D^2}{\rho_{tot}}$.}\\
 The dynamical equations (\ref{deqs}) in the presnt case takes to the form
\bea
x'&=&\frac{3 b^2 \left(x^2+y^2-1\right)^2 \left(2 m^2+2 x^2+y^2-2\right)+x^2 \left(2 m^2+2 x^2+5 y^2-2\right)}{2 x \left(2 m^2+y^2-2\right)}\,,\nn\\
y'&=&\frac{y \left(3 b^2 \left(x^2+y^2-1\right)^2+4 m^2+x^2+4 y^2-4\right)}{2 \left(2 m^2+y^2-2\right)}\,.
\eea
Investigating these equations provides just two critical points. Similar to the interaction II, the $x=0$ line is excluded form the phase space since the dynamical equation $x'$ diverges on it. The critical points are

\begin{figure}[tt]
\begin{picture}(0,0)(0,0)
\put(80,-12){\footnotesize Fig (3.a)} \put(310,-12){\footnotesize Fig (3.b)}
\put(49,-185){\footnotesize Fig (3.c)}
\put(210,-185){\footnotesize Fig (3.d)}
\put(392,-185){\footnotesize Fig (3.e)}
\put(35,-30){\footnotesize L}
\put(36,-39){$\bullet$}
\end{picture}
\qquad \includegraphics[height=50mm,width=48mm,angle=0]{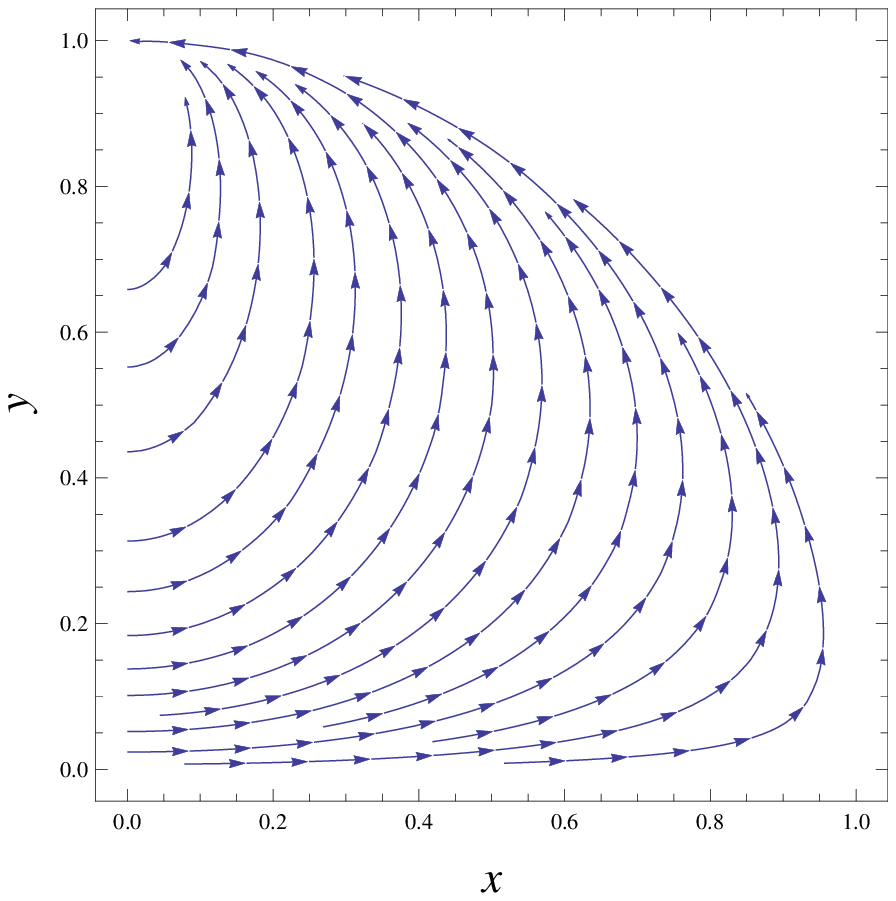} \hspace{2cm} \includegraphics[height=50mm,width=68mm,angle=0]{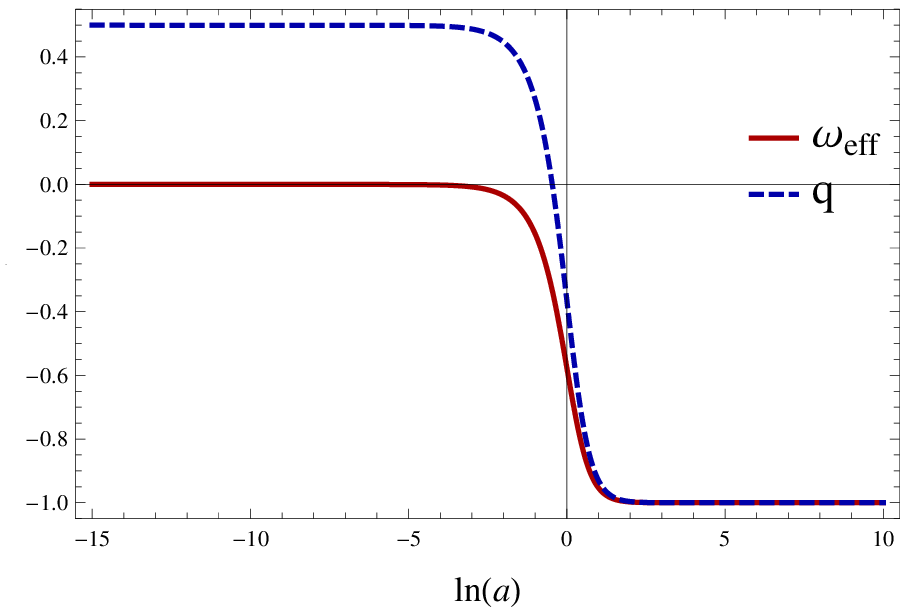}\\

\vspace{8mm}
\!\!\!\!\!\includegraphics[height=50mm,width=46mm,angle=0]{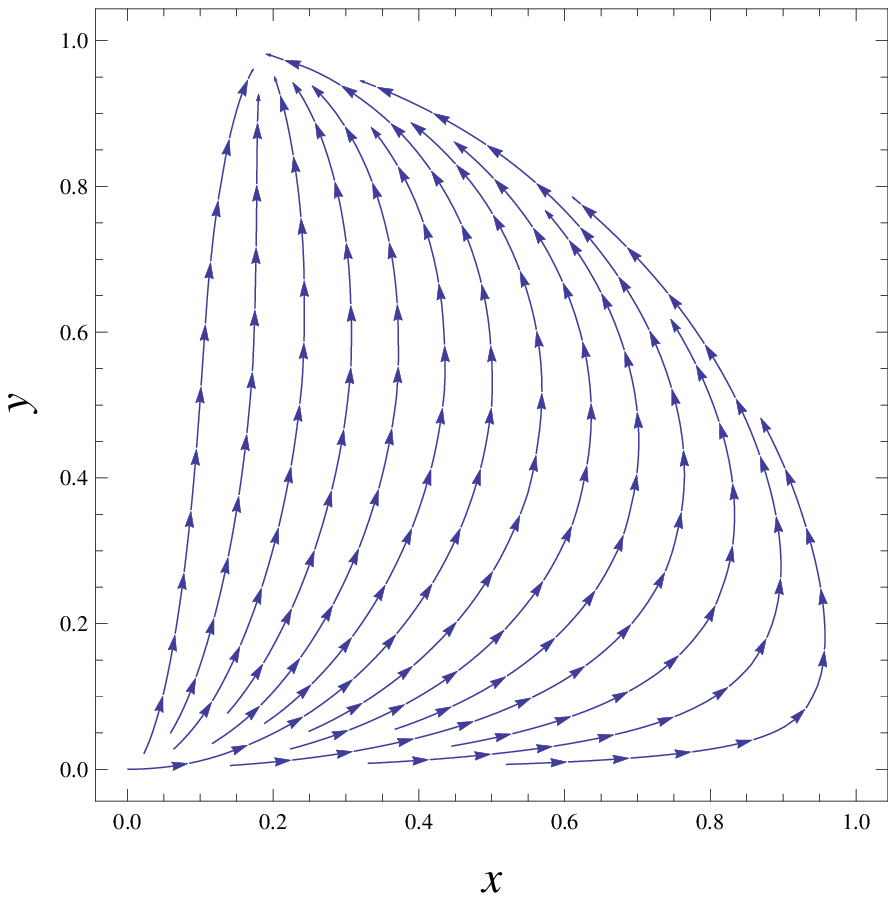}
\,\,\includegraphics[height=48mm,width=62mm,angle=0]{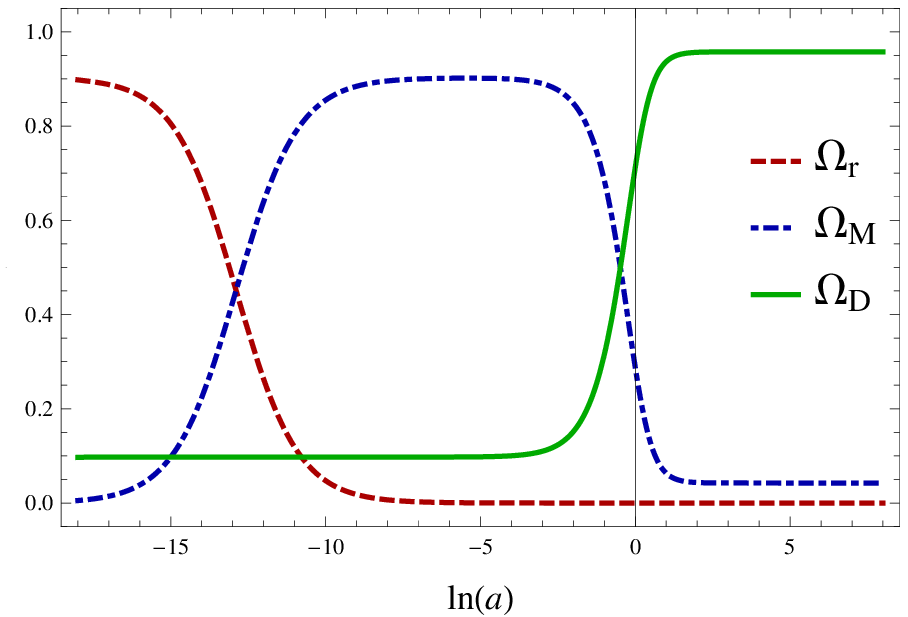}\,\,\includegraphics[height=48mm,width=60mm,angle=0]{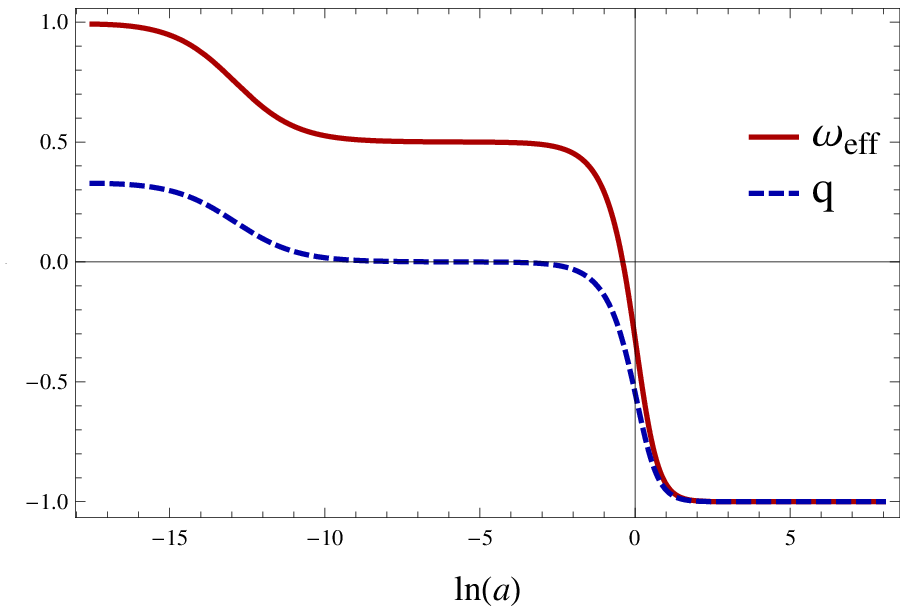}

\vspace{4mm} \caption{{\footnotesize (3.a), (3.b) shows the evolution of phase space and cosmological parameters for GDE with non-linear interactions V, by choosing $b=0.2$\,. The model does not show radiation dominated phase.
The radiation dominated phase is recovered in the case of non-linear interaction VI and so the GDE model shows a normal behavior with interaction VI as it is depicted in (3.c), (3.d) and (3.e) by choosing $b=0.2$\,. The late time attractor point (L) in this case is a dark energy/matter scaling phase with ratio $\Omega_D/\Omega_m\sim 96/4=24$\,.}}
 \label{fig6}
\end{figure}

$\bullet$ \, $P_1$: $(x=\sqrt{1-m^2},\, y=0)$, \, shows a
matter/EDE scaling phase in the universe with $\Omega_m\sim 0.9$. In this case one can obtains $q=\frac 12+\frac{3b^2 m^4}{2(m^2-1)}\sim \frac 12$ and $w_{eff}=-b^2 m^4\sim 0$ (figure \ref{fig6}.b). The
instability of the matter dominated epoch is obvious from
the eigenvalues of the stability matrix
$\lambda_1=\frac 34$ and $\lambda_2=-1$\,.

$\bullet$ \, $P_2$: $(x=bm^2,\,y=\sqrt{1-m^2-b^2m^4}\,)$. \, Since $m^2$ in negligible at late times, this critical point
describes a dark energy dominated phase of the universe . $P_2$ is stable due to the negative eigenvalues $\lambda_1=-4$ and $\lambda_2=-3$\,.  For this
fixed point one finds the deceleration and EoS parameters as
$q=-1$ and $w_D=w_{eff}=-1$\,.

 Since the absence of $x=0=y$ critical point (radiation dominated epoch in the early times) the GDE model with non-linear interaction V is not physically accepted. Figures (\ref{fig6}.a) demonstrate the phase plane of GDE with non-linear interactions V.  The evolution of $w_{eff}$ and $q$ in this case depicted in figure (\ref{fig6}.b). 

\vspace{3mm}
{\bf VI) The non-linear interaction $Q=3b^2 H \frac{\rho_D^3}{\rho_{tot}^2}$\,.}\\
 Finally in this case the dynamical equations are
\bea \label{deq6}
x'&=&\frac{3 b^2 \left(m^2+y^2\right)^3 \left(2 \left(m^2+x^2-1\right)+y^2\right)+x^2 \left(2 \left(m^2+x^2-1\right)+5 y^2\right)}{2 x \left(2 m^2+y^2-2\right)}\,, \nn\\
y'&=&\frac{y \left(3 b^2 m^6+9 b^2 m^4 y^2+m^2 \left(9 b^2 y^4+4\right)+3 b^2 y^6+x^2+4 y^2-4\right)}{2 \left(2 m^2+y^2-2\right)}\,.
\eea
 It seems that $x'$ in the above diverges at $x=0$, however in the limit of $y\to 0$ one finds that
\be \label{x'6}
x'=\frac{\left(m^2+x^2-1\right) \left(3 b^2 m^6+x^2\right)}{2 \left(m^2-1\right) x}\,,
\ee
considering the values of parameters $b=0.2$ and $ 0<m \leq 0.1$, one finds that $3b^2m^6\approx 10^{-7}$ so it is possible to ignore this value even in the case of small matter density $\left(\Omega_m=x^2\right)$\,. In the other words (\ref{x'6}) remains smooth in the case of $x\approx 0 \approx y$\,. On the other hand, by ignoring $3b^2m^6$ in (\ref{x'6}) one finds $x' \propto x$ which means that $x\approx 0 \approx y$ is a physical fixed point.
Hence the dynamical equations (\ref{deq6}) shows three physical fixed points as

$\bullet$ \, $P_1$: $(x\approx 0,\, y\approx 0)$. Unlike the previous interaction, the radiation dominated critical point (with $\Omega_r\sim 0.9$) reappears in the phase space of the model. In fact $P_1$ describes an unstable\footnote{The eigenvalues of the stability matrix are messy to be written here, but we checked that one of them is positive, so the phase is unstable.} radiation phase where $q=1$ and $w_{eff}=\frac 13$\,.

$\bullet$ \, $P_2$: $(x=\sqrt{1-m^2},\, y=0)$, \, which is according to a
matter/EDE scaling epoch at the early universe. Using (\ref{wqw}) one finds $q=\frac 12+\frac{3b^2 m^6}{2(m^2-1)}\sim \frac 12$ and $w_{eff}=-b^2 m^6\sim 0$ (figure \ref{fig6}.e). This matter dominated phase is unstable due to the eigenvalues $\lambda_1\sim \frac 34$ and $\lambda_2\sim -1$\,.

The critical point $P_3$ is somewhat messy to show it here, but we checked that it corresponds to a stable 
dark energy/matter scaling phase in the late time universe. 

The phase plane of GDE wit non-linear interaction VI is depicted in figure (\ref{fig6}.c). It is clear that there are radiation/EDE, matter/EDE and dark energy/matter scaling fixed point in this model. Figure (\ref{fig6}.d) shows the evolution of fractional energy densities. Tunning the initial conditions we found $\Omega_D \approx 0.7$ and $\Omega_m \approx 0.3$ at present time ($z=0=\ln a$). In this model we encounter early dark energy in the past times, with constant fractional energy $\Omega_{EDE}\sim 0.1$\,. Hence the model starts at radiation/EDE scaling phase with $\Omega_r\sim 0.9$, then it passes a matter/EDE ($\Omega_m\sim 0.9$) epoch, and finally reaches a stable dark energy/matter scaling phase where $\Omega_D=1-b^2$. We also plotted the evolution of cosmological parameters $w_{eff}$, $q$ in figure (\ref{fig6}.e)\,. 

\section{The evolution of interacting HDE} \label{dshde}
In this section we consider the phase space analysis of HDE model. 
Similar to the previous section We start from non-interacting HDE, then we add the linear
interaction between dark matter and dark energy, and finally we investigate the phase space of the model with non-linear
interactions. In this case,
supposing an interaction term $Q$ between DM and the DE components,
the continuity equations take to the form (\ref{conti1}) and (\ref{conti2}). By differentiating (\ref{hdeden}) and using (\ref{conti2}) one finds 
\be \label{hde1}
3H\rho_D(1+w_D)+Q=2\dot{R_h}{R_h}^{-1}\,\rho_D\,,
\ee
noticing that $\dot{R_h}=HR-1$, the EoS parameter for dark energy could be obtained easily by solving the above equation. It is also possible to find the deceleration parameter. The result is
\bea \label{hdewq}
w_D&=&-\frac{1}{9} \left(\frac{6 \sqrt{\text{$\Omega_D $}}}{c}+\frac{8 \pi  G Q}{H^3 \text{$\Omega_D$}}+3\right)\,, \nn \\
q&=&-1-\frac{\dot H}{H^2}=1-y^2-\frac{y^3}{c}-\frac{x^2}{2}-\frac{4 \pi  G Q}{3 H^3}\,,
\eea
where we use $R_h=\frac{c}{H\sqrt{\Omega_D}}$\,. We have also introduced the dynamical variables $x^2=\Omega_m$ and $y^2=\Omega_D$ as in (\ref{xydef}). Note that the radiation density parameter is not independent variable, it satisfies $\Omega_r=1-x^2-y^2$. Now the dynamical equations take to the form
\bea \label{dyneqs}
x'&=&\frac{x}{2}-\frac{x^3}{2}-x y^2-\frac{x y^3}{c}-\frac{x}{2}\,g(x,y)+\frac{g(x,y)}{2 x}\,, \nn\\
y'&=&y\left[1-\frac{x^2}{2}+\frac{y}{c}-y^2-\frac{y^3}{c}-\frac{1}{2}\,g(x,y)\right],
\eea
where the prime denotes derivation with respect to $\ln a$, we also introduce $g(x,y)=\frac{\Omega_q}{H}=\frac{8\pi G Q}{3H^3}$.  As we will show, because of the presence of $x$ in $g(x,y)$\,, the dynamical equations of HDE model with interactions III, IV\, are smooth in the range of variables. In the case of interactions V, VI the dynamical equations are conditionally smooth at $x=0$, However for the linear interaction, $x=0$ is a singularity. In this case, similar to GDE, considering the interaction between DM and DE as (\ref{genint}), one deduces that the phase space of the model is two dimensional. In the following we will investigate the fixed points  of the dynamical equations (\ref{dyneqs}) for different interactions mentioned before and discuss the evolution of  the relevant interacting HDE model.

\vspace{3mm}
{\bf I) The case of\, $Q=0$\,.}

In the non-interacting HDE model,  setting $g(x,y)=0$ in (\ref{dyneqs}), it is obvious that the dynamical equations are smooth and one finds three fixed points:

$\bullet$ \, $F_1$: $(x=0,\, y=0)$. \, This point corresponds to
the radiation dominated phase of the non-interacting HDE model. In this case using (\ref{teos}) and (\ref{hdewq}) one finds that $w_{eff}=\frac13$ and $q=1$ respectively, which means the universe is decelerating in this phase. The eigenvalues of the stability matrix are $\lambda_1=1/2$ and $\lambda_2=1$ that shows the non interacting HDE model provide unstable radiation dominated phase.

$\bullet$ \, $F_2$: $(x=1,\, y=0)$. \, The matter-dominated epoch is described by this critical point. The eigenvalues of stability matrix in this case are $\lambda_1=-1$, $\lambda_2=1/2$ and using (\ref{teos}), (\ref{hdewq}) one finds $w_{eff}=0$, $q=1/2$\,. Therefore, this phase is an unstable and unaccelerated,  as one expects.  

$\bullet$ \, $F_3$: $(x=0,\, y=1)$. \, This point shows the  dark energy dominated phase of HDE model. At this stage using (\ref{hdewq}) one finds $q=-\frac{1}{c}$ and $w_D=-\frac 13(1+\frac 2c)$\,. Remember that $c>0$ so $w_D<-1/3$ is guaranteed and $q$ is negative. Note also that in the case of $c=1$, the non-interacting HDE behaves as $\Lambda$CDM and for $c<1$ the model shows the phantom behavior. In this phase, the stability matrix has eigenvalues $\lambda_1=-\frac{2+2c}{c}$ and $\lambda_2=-\frac{2+c}{2c}$ which both are negative and this confirms that the dark energy dominated phase of the model is stable. For this solution of dynamical system, $w_{eff}=-1$ which is consistent with $\Lambda$CDM model.

\vspace{3mm}
{\bf II) The case of\, $Q=3b^2 H\rho_{tot}$.}\\
The dynamical equations (\ref{dyneqs}) in this case takes to the form
\be
x'\!=-\frac{3 b^2 x}{2}+\frac{3 b^2}{2 x}-\frac{x^3}{2}-x y^3-x y^2+\frac{x}{2}, \quad y'\!=-\frac{y}{2} \left[3 b^2\!+x^2\!+2 (y-1) (y+1)^2\right].
\ee
 These equations reveal the deficiency of HDE with linear interaction: $x'$ is singular at $x=0$, hence there is no radiation dominated fixed point in the model. In fact the dynamical equations possess just two fixed points:

$\bullet$ \, $F_1$: $(x=1,\, y=0)$.  The instability of this matter dominated phase is obvious from the Eigenvalues of stability matrix: $\lambda_1=\frac{1}{2} \left(1-3 b^2\right)$ and $\lambda_2=- \left(1+3 b^2\right)$ where the coupling constant $b$ is a very small positive number ($b<\frac{1}{\sqrt{3}}$). The deceleration and EoS parameters in this phase $q=\frac 12(1-3b^2)$ and $w_{eff}=-b^2$\,, indicate that matter dominated era is decelerating due to small value of $b$.

While the second fixed point (which is too messy to be written here) describe a stable dark energy-matter scaling phase, the model suffers the absence of radiation dominated epoch. In fact, the linear interaction in the context of both dark energy models, GDE and HDE, does not provide true cosmological consequences of expected eras.

\begin{figure}[tt]
\begin{picture}(0,0)(0,0)
\put(102,0){\footnotesize Fig (4.a)} \put(340,0){\footnotesize
Fig (4.b)}
\end{picture}
\center
\includegraphics[height=70mm,width=72mm,angle=0]{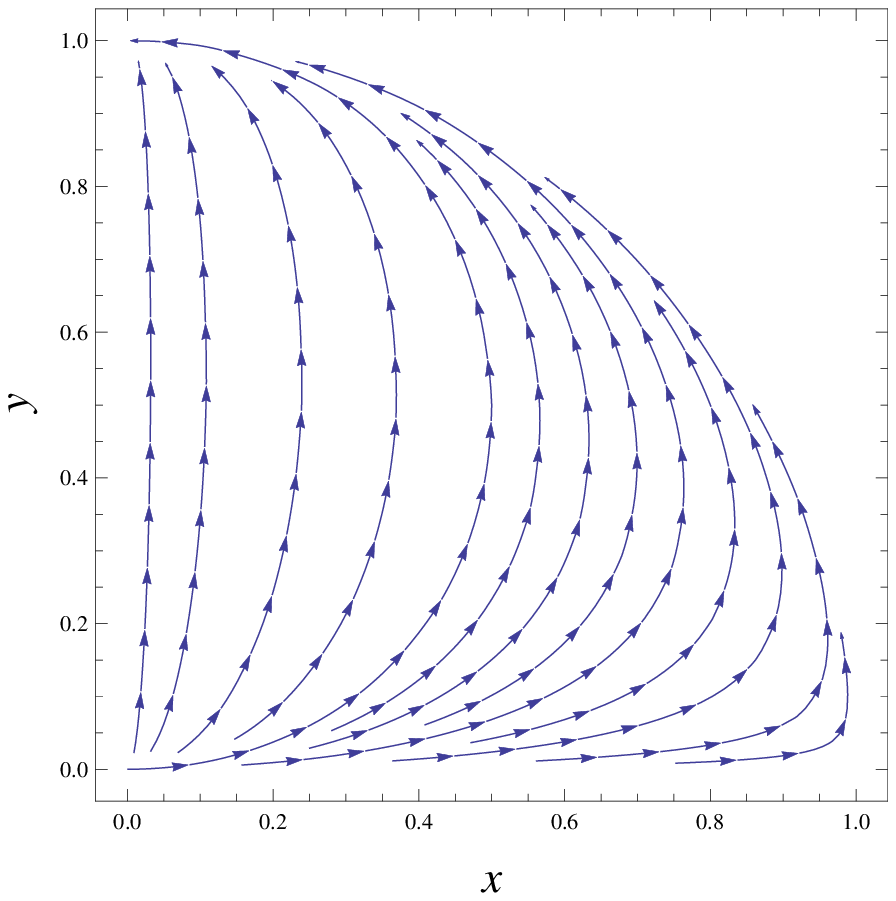} \quad
\includegraphics[height=70mm,width=72mm,angle=0]{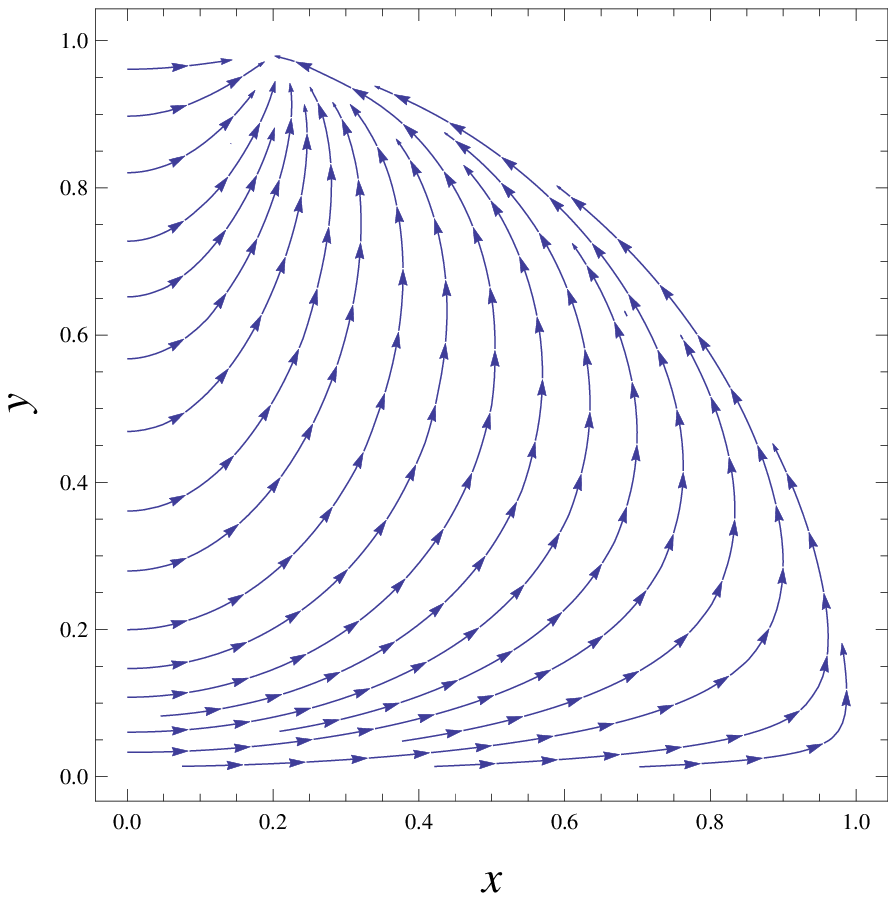}
 \caption{{\footnotesize The evolution of phase space
by choosing $b=0.2, c=1$\, for non-interacting HDE (4.a) and HDE with linear interaction (4.b). There
is unstable radiation dominated phase, unstable matter dominated
epoch and stable HDE dominated in non-interacting HDE but  the radiation dominated phase destroys by adding the linear interaction. This failure resolved by adding the non-linear terms.}} \label{fig7}
\end{figure}

 We will show in the following that this failure improved when we add the non-linear interaction terms to the HDE.  Figure (\ref{fig7}) shows the evolution of universe containing radiation, matter, DM and HDE in the absence of interaction (Fig \ref{fig7}.a) and in the presence of the linear interaction (Fig \ref{fig7}.b). It is obvious from Fig(\ref{fig7}.a) that all arrows ends at ($x=0$, $y=1$) which is late time attractor of HDE dominated universe. In true cosmological paths the universe starts from unstable radiation dominated phase, passes a transient matter dominated phase and finally reach a stable HDE dominated phase. At first glance to Fig(\ref{fig7}.b) it may seems that there is a fixed line $x=0$ and true cosmic paths could starts from ($x=0$, $y=0$); but note that setting $x=0$ in (\ref{dyneqs}), $x'$ diverges and so there is no radiation dominated fixed point in the HDE model with linear interaction.

\vspace{3mm}
{\bf III) \,  The non-linear interaction $Q=3b^2 H\frac{\rho_D \rho_m}{\rho_{tot}}$\,.}\\
By adding the above interaction to the HDE model, one finds that dynamical equations (\ref{dyneqs}) can be rewritten as
\bea
&&x'=\frac{x}{2c} \left[c y^2 \left(-3 b^2 \left(x^2-1\right)-2\right)-c x^2+c-2 y^3\right], \nn\\ &&y'=\frac{y}{2c} \left[-c \left(y^2 \left(3 b^2 x^2+2\right)+x^2-2\right)-2 y^3+2 y\right],
\eea
which are smooth everywhere. They have three physical fixed points:

$\bullet$ \, $F_1$: $(x=0,\, y=0)$. This critical point describes the expected radiation dominated phase in the evolution of the universe. because of the positivity of the eigenvalues $\lambda_1=1/2$ and $\lambda_2=1$, this phase is unstable. In this point using (\ref{teos}), (\ref{hdewq}) one can obtain $w_{eff}=\frac13$\,, $q=1$ which shows the deceleration of a standard radiation dominated universe.

$\bullet$ \, $F_2$: $(x=1,\, y=0)$. Unstable matter dominated phase of the universe describes by this fixed point where the eigenvalues are $\lambda_1=1/2$ and $\lambda_2=-1$. The deceleration and EoS parameters for this phase could be found as $q=1/2$\,, $w_{eff}=0$\,.

$\bullet$ \, $F_3$: $(x=0,\, y=1)$. This is the late time attractor of DE dominated universe. Using (\ref{hdewq}) one finds $w_D=-\frac 13(1+\frac{2}{c})$ and $q=-\frac{1}{c}$ which, noticing $c>0$, present the accelerating evolution of the universe. Note also that in the case of $c<1$, HDE  shows the phantom behavior. This phase is stable due to the negativity\footnote{Note that $b^2$ is so small and $c>0$.} of the eigenvalues $\lambda_1=-\frac{2 c+2}{c}$, $\lambda_2=\frac{3 b^2c -c-2}{2 c}$\,. It is also easy to find that $w_{eff}=-1$ in this epoch.

\vspace{3mm}
{\bf IV) \,  The non-linear interaction $Q=3b^2 H\frac{\rho_m^2}{\rho_{tot}}$.}\\
Similar to the previous case, substituting the above interaction in (\ref{dyneqs}) one finds that the dynamical equations are smooth and there are three acceptable fixed points as

\vspace{1mm}
$\bullet$ \, $F_1$: $(x=0,\, y=0)$. As we mentioned, the radiation dominated phase is recovered for HDE by adding the non-linear interaction terms. The instability of this period is obvious form the eigenvalues of the stability matrix $\lambda_1=1$ and $\lambda_2=1/2$. In this case, one finds $q=1$\,, $w_{eff}=\frac13$\, which shows the decelerating feature of the radiation dominated phase.

$\bullet$ \, $F_2$: $(x=1,\, y=0)$, corresponds to an unstable matter dominated phase in the universe, since the eigenvalues are $\lambda_1=\frac{1}{2} \left(1-3 b^2\right)$ and $\lambda_2=\frac{1}{2} \left(1-3 \left(2 b^2+1\right)\right)$. In this case, using (\ref{hdewq}), (\ref{teos}) one also finds $q=\frac{1}{2} \left(1-3 b^2\right)$\,, $w_{eff}=-b^2$\,. Note that the positivity of $\lambda_1$ and $q$ and also the necessity of $w_{eff}>-\frac13$ in this epoch, put an upper bound on the coupling constant of the interactions of DE and DM as $b<0.57$; in the other words, this matter dominated phase is unstable and decelerating if $b<0.57$\,.

$\bullet$ \, $F_3$: $(x=0,\, y=1)$. The dark energy dominated phase is described by this fixed point, where one can read from (\ref{hdewq}) that $w_D=-\frac 13(1+\frac{2}{c})$ and $q=-\frac{1}{c}$\,. So the HDE shows phantom behavior when $c<1$\,. The DE dominated phase is stable because of the negative eigenvalues $\lambda_1=-\frac{2+2c}{c}$ and $\lambda_2=-\frac{2+c}{2c}$\,.

\begin{figure}[tt]
\begin{picture}(0,0)(0,0)
\put(4,135){\footnotesize L}
\put(5,125){$\bullet$}
\put(43,-15){\footnotesize Fig (5.a)} \put(205,-15){\footnotesize
Fig (5.b)}
\put(392,-15){\footnotesize Fig (5.c)}\end{picture}
\!\!\!\!\!\!\!\!\includegraphics[height=47mm,width=45mm,angle=0]{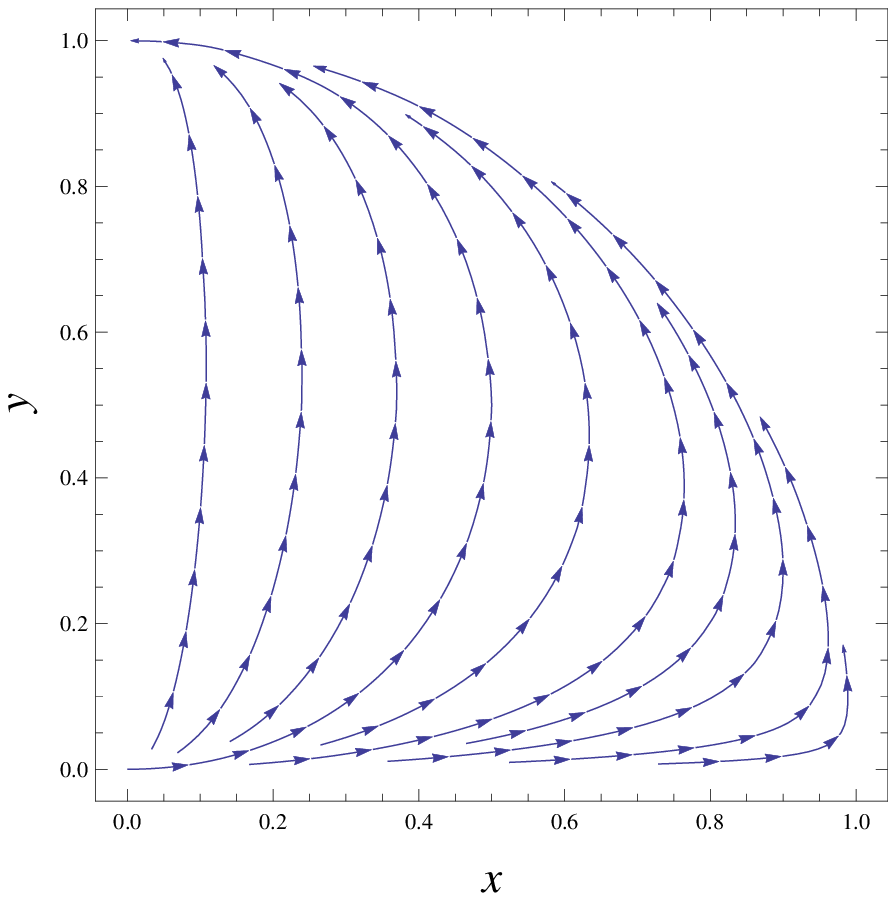}\,\,\,\,\includegraphics[height=47mm,width=65mm,angle=0]{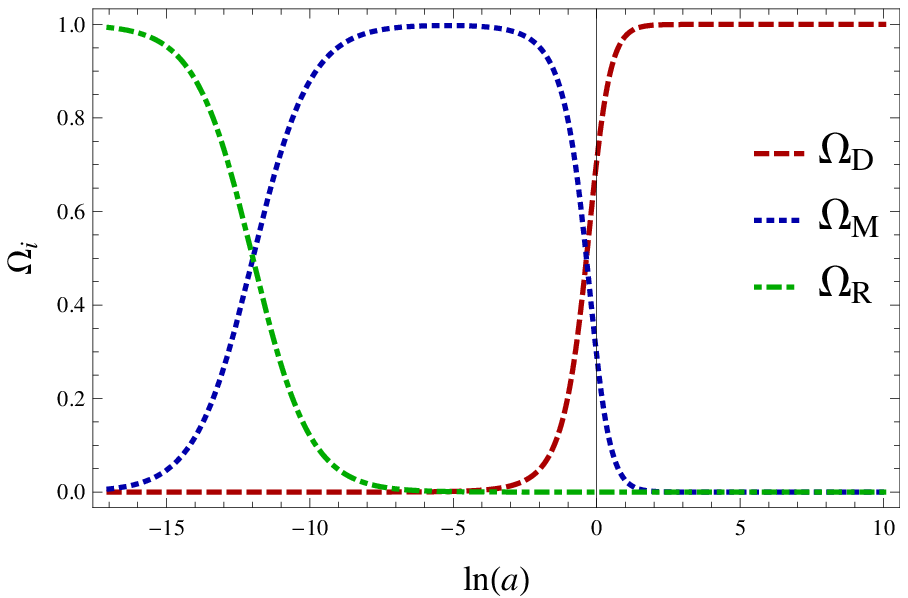}\,\,\,\,\includegraphics[height=47mm,width=60mm,angle=0]{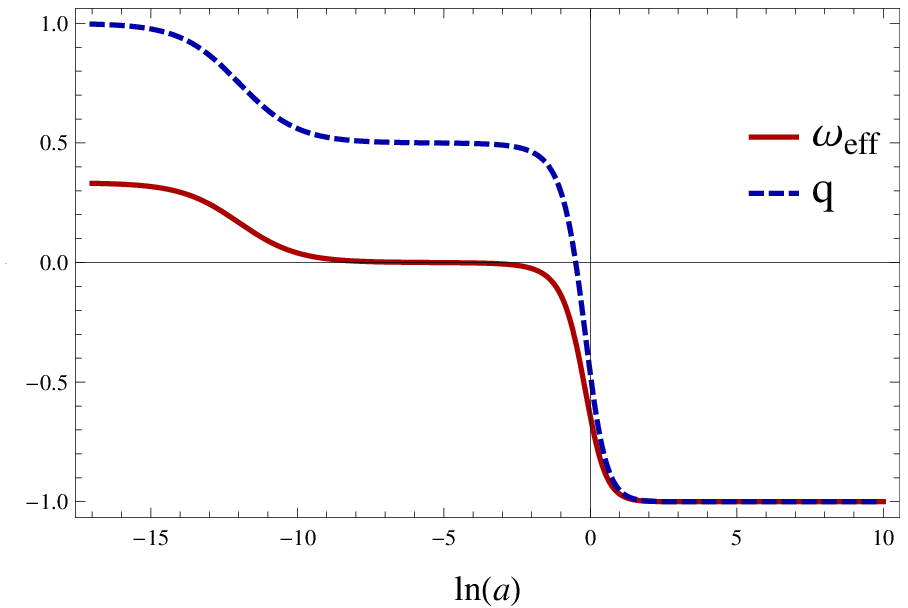}
\vspace{2mm} \caption{{\footnotesize (5.a) Shows the evolution of phase space for non-linear interactions III and IV by choosing $b=0.2,\, c=1$\,. Note that the radiation dominated fixed point (which was absent in the case of linear interaction) is recovered, so the models contain unstable radiation dominated phase, unstable matter dominated
epoch and stable HDE dominated phase (the attractor point $L$).
\, In (5.b), by choosing adequate initial conditions, we depicted the evolution of density parameters.\, (5.c) shows the variations of EoS and deceleration parameters.}} \label{fig8}
\end{figure}

We plotted the phase space of HDE with non-linear interactions III and IV. The evolution of phase space  is very similar in these two cases and it has been presented in figure (\ref{fig8}.a). It is easy to see that there is a late time attractor point $L:(x=0, y=1)$ in correspondence with DE dominated phase in the universe. As a main result, we see that the radiation dominated fixed point ($x=0, y=0$) is recovered by adding the non-linear interaction terms. Remember that this point was absent in the case of HDE with linear interaction. In figure (\ref{fig8}.a) there are different paths corresponding to different initial conditions at the early universe, however, true cosmic evolution belongs to the paths which started at unstable radiation dominated  fixed point ($x=0=y$), pass through the unstable matter dominated phase ($x\approx 1, y\approx 0$), reach the region $x^2\approx 0.3$, $y^2\approx 0.7$ at present finally end at stable dark energy dominated fixed point $L$\,.

Figure (\ref{fig8}.b), also shows the evolution of density parameters for the components of universe, in the case of non-linear interactions III, IV. Noting that $\ln a=-\ln (1+z)$, it is easy to see that the energy density of matter becomes dominant after the early stages of the universe and there is another phase transition between matter and dark energy, near the present time. By setting the initial conditions, one can find the present day values $\Omega_D \approx 0.7$ and $\Omega_m \approx 0.3$ when $\ln a=0=z$\,. 
The evolution of EoS and deceleration parameters also depicted in figure (\ref{fig8}.c).

\vspace{3mm}
{\bf V) \,  The non-linear interaction $Q=3b^2 H\frac{\rho_D^2}{\rho_{tot}}$.}\\
For this type of interaction the dynamical equations (\ref{dyneqs}) take to the form
\bea
&&x'=-\frac{1}{2} 3 b^2 x y^4+\frac{3 b^2 y^4}{2 x}-\frac{x y^3}{c}-\frac{x^3}{2}-x y^2+\frac{x}{2}\,, \nn\\
&&y'=-\frac{y}{2c} \left[c \left(3 b^2 y^4+x^2+2 y^2-2\right)+2 y \left(y^2-1\right)\right]\,.
\eea
 It seems that $x'$ in the above diverges at $x=0$, however $x'$ on the $x=0$ line is well behaved in the limit $y\to 0$ so have three acceptable fixed points 

$\bullet$ \, $F_1$: $(x=0,\, y=0)$. The unstable radiation dominated phase, described by this fixed point where the eigenvalues of the stability matrix are $\lambda_1=1/2$ and $\lambda_2=1$. For this phase, Using (\ref{teos}), (\ref{hdewq}) one can find that $w_{eff}=\frac13$\,, $q=1$\,.

$\bullet$ \, $F_2$: $(x=1,\, y=0)$. This is the unstable matter dominated phase because of the eigenvalues $\lambda_1=1/2$ and $\lambda_2=-1$. The deceleration parameter in this case is $q=1/2$ and also $w_{eff}=0$\,.This point corresponds to a standard matter dominated epoch.

\begin{figure}[tt]
\begin{picture}(0,0)(0,0)
\put(23,132){\footnotesize L}
\put(21,124){$\bullet$}
\put(43,-15){\footnotesize Fig (6.a)} \put(205,-15){\footnotesize
Fig (6.b)}
\put(392,-15){\footnotesize Fig (6.c)}
\end{picture}
\!\!\!\!\!\!\!\!\includegraphics[height=47mm,width=45mm,angle=0]{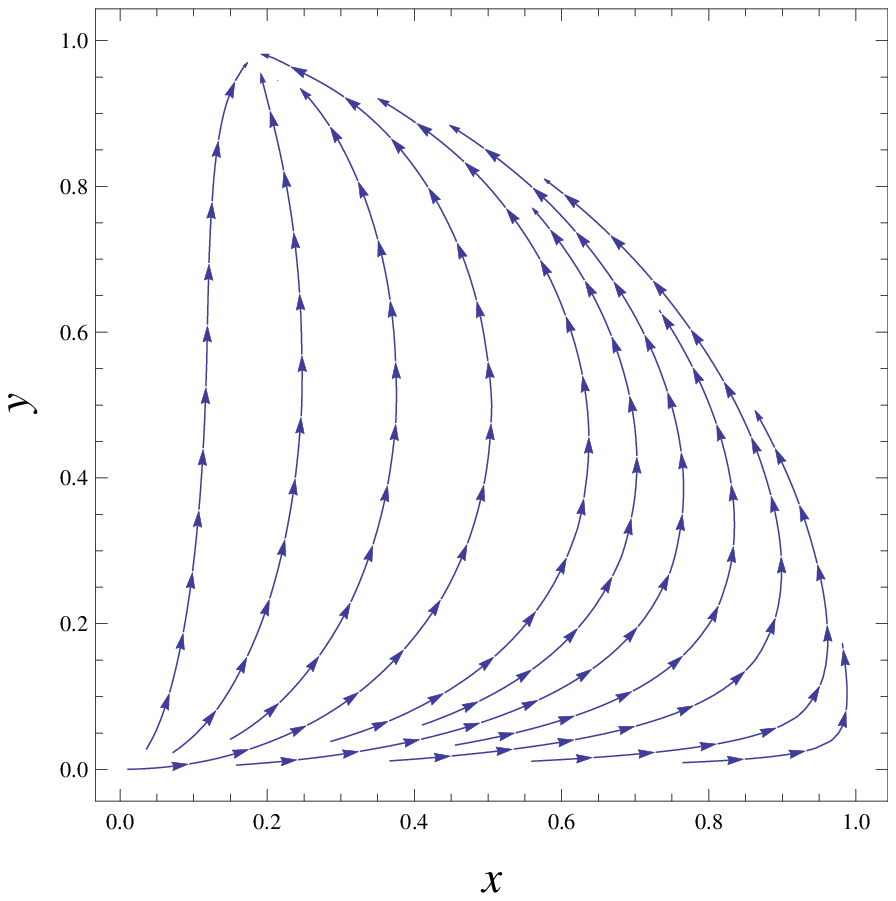}\,\,\,\includegraphics[height=47mm,width=63mm,angle=0]{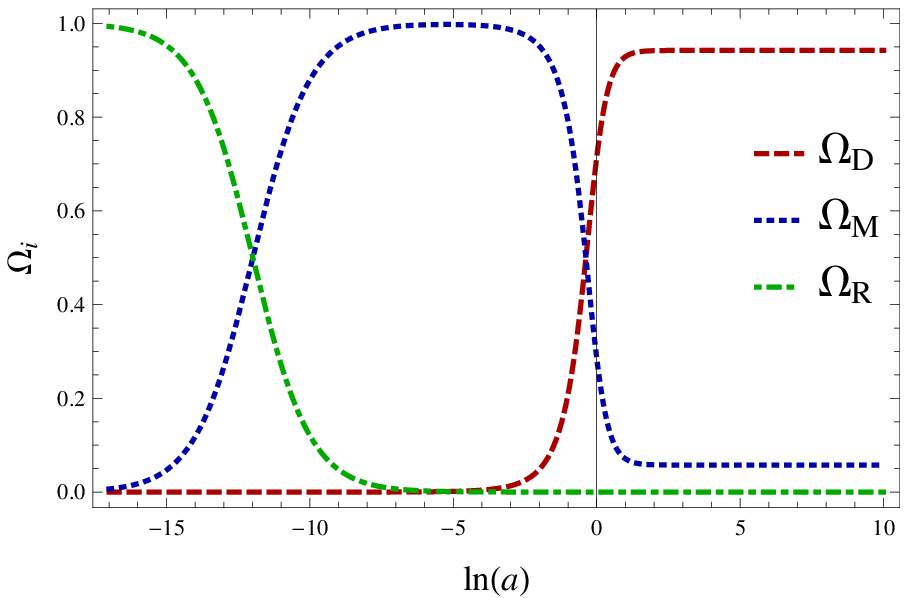}\,\,\,\includegraphics[height=47mm,width=60mm,angle=0]{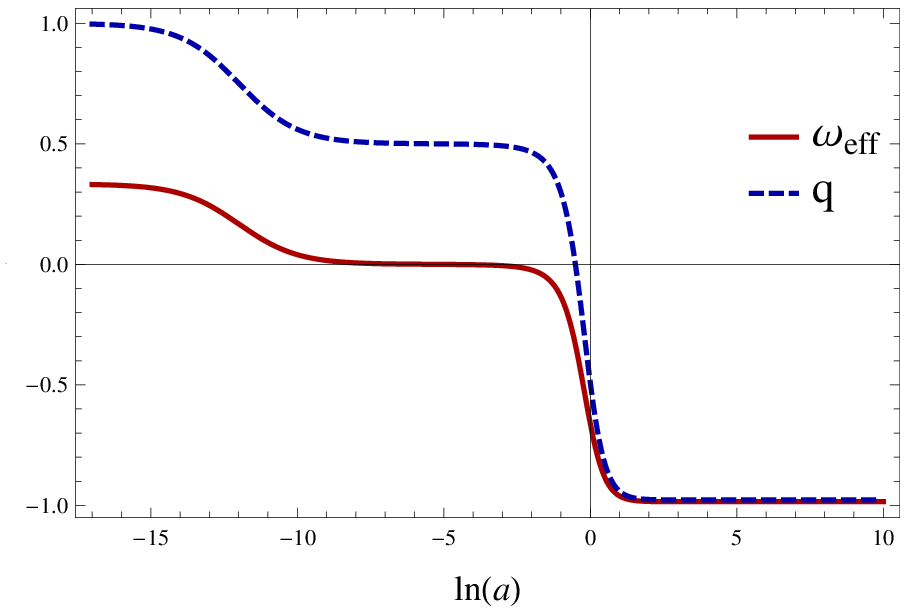}
\vspace{2mm} \caption{{\footnotesize The evolution of phase space for non-linear interactions V, VI plotted in (6.a)
by choosing $b=0.2, c=1$\,.  In both cases, there
is unstable radiation dominated phase, unstable matter dominated
epoch and stable dark energy-matter scaling phase. Choosing  the proper initial conditions, density parameters evolve as figure (6.b). Note that at the late times, as a scaling solution, the ratio of $\Omega_D/ \Omega_m \approx 19$\,.}} \label{fig9}
\end{figure}

The third fixed point of this model is too messy, but we checked that it demonstrate a stable dark energy-matter scaling phase in the evolution of the universe, as shown in figure \ref{fig9}\,.

\vspace{3mm}
{\bf VI) \,  The non-linear interaction $Q=3b^2 H\frac{\rho_D^3}{\rho^2_{tot}}$.}\\
Finally in this case, the dynamical equations can be found as, 
\bea
&&x'=-\frac{1}{2} 3 b^2 x y^6+\frac{3 b^2 y^6}{2 x}-\frac{x y^3}{c}-\frac{x^3}{2}-x y^2+\frac{x}{2}\,, \nn \\
&&y'=-\frac{y}{2c} \left[c \left(3 b^2 y^6+x^2+2 y^2-2\right)+2 y \left(y^2-1\right)\right].
\eea
 Similar to the previous case, the singularity of $x'$ removes at the limit of $y\to 0$ and the dynamical equations accept three physical fixed points

$\bullet$ \, $F_1$: $(x=0,\, y=0)$. Similar to the previous cases this point describes unstable radiation dominated phase in which $q=1$\, and $w_{eff}=\frac13$\,.

$\bullet$ \, $F_2$: $(x=1,\, y=0)$\,, corresponds to a matter dominated phase in the evolution of the universe with $w_{eff}=0$\,. The phase is unstable due to the eigenvalues $\lambda_1=1/2$ and $\lambda_2=-1$. It is also decelerating since $q=1/2$\,.

The third fixed point which is so messy to written here, shows a stable dark energy-matter scaling phase in the late time evolution of the model. We plotted the behavior of the system in figure \ref{fig9}\,.

Note that similar to the other non-linear interactions mentioned before, the radiation dominated phase is recovered by adding non-linear terms V, VI to the interacting HDE models. In figure (\ref{fig9}.a) we have depicted the phase space evolution of HDE with non-linear interactions V and VI, which contains unstable radiation dominated phase, unstable matter dominated one and stable matter-DE scaling phase. Figure (\ref{fig9}.b) also shows the evolution of fractional density parameters. By tunning the initial conditions we found $\Omega_D \approx 0.7$ and $\Omega_m \approx 0.3$ at present. Note that at late times we encounter to a DE-matter scaling phase which $\frac{\Omega_D}{\Omega_m}\approx \frac{0.95}{0.5}=19$\,. The variations of EoS and deceleration parameters is depicted in figure (\ref{fig9}.c).

\section{Discussion and conclusion}
In this paper using the dynamical system analysis, we studied the impacts of interaction between dark
energy and dark matter on the evolution of the universe in the context of two dark energy models: ghost dark energy and holographic dark energy.
In the absence of interaction, both models shows normal behavior: they 
start from radiation dominated epoch in the early times, pass the unstable matter dominated era and finally reach to stable
dark energy epoch. Note that in the case of GDE, there is a constant part $m^2$ in $\Omega_D$ which could play role in the early evolution of the universe and according to \cite{caighost2} it has around $10\%$ of fractional energy density in the form of early dark energy (EDE). In other words, in the radiation dominated phase of the GDE model we have $\Omega_r\sim0.9$ and $\Omega_{EDE}\sim 0.1$\,.

We then add the linear interaction between dark
matter and dark energy. In this case, the dynamical equations of both models do not contain fixed point around $(x=0, y=0)$, in other
words, GDE and HDE accompanied by the linear interaction do not have the radiation dominated epoch in the early times so they are not physically accepted. This
failure improved when we replace linear interaction with the non-linear one.

We also discussed the evolution of GDE and HDE models containing  non-linear
interactions. In the case of HDE we found that the radiation dominated phase is recovered by adding  non-linear interactions and the model starts normally from
radiation dominated era, pass through unstable matter dominated
epoch and finally end at dark energy dominated (for interactions III, IV) or dark
energy-matter scaling epoch with $\Omega_D/\Omega_m \approx 19$ (for interactions V, VI) 
in the late time. In the other words, HDE model with the non-linear interactions is cosmologically accepted.

 In the case of GDE model, we found that the radiation dominated phase, is recovered by adding non-linear interactions III, IV and VI to the GDE. In the case of interactions III and IV, the model starts from an unstable radiation dominated epoch with $\Omega_D \approx 0.9$, passes an unstable matter dominated era and finally reaches a stable dark energy dominated phase. In the presence of interaction VI, GDE shows the same radiation and matter dominated epochs but the final state is a stable dark energy/matter scaling phase with $\Omega_D /\Omega_m \approx 24$. However, addition of a non-linear interaction term in the form V can not recover the radiation dominated phase; so the GDE model with interaction V is not physically accepted.

Generically the above results are independent from the value of coupling constant $b$ for  the interactions between dark energy and dark matter. In fact just for the interaction IV the value of $b$ has a bound were we found that there is an upper bound as $b<0.61$ in the case of GDE. We also found that there is a similar bound in the case of HDE model as\, $b<0.57$\,. These bounds come from two facts that the matter dominated phase of the universe should be unstable (the eigenvalues of the stability matrix should be negative) and also the acceleration of the universe should be negative in the matter dominated epoch.


\section*{Acknowledgment}
We would like to thank Dr. Mahmood Roshan for his useful comments. The work of Esmaeil Ebrahimi has been supported by Research Institute for Astronomy
and Astrophysics of Maragha (RIAAM), Iran.



 \end{document}